\documentclass[twocolumn,showpacs,pra]{revtex4}
\usepackage{graphicx}
\usepackage{amssymb}
\begin{document}
%\jl{2}    %JPB
%\draft    %JPB

\title{Tree-body loss of of trapped ultracold $^{87}$Rb atoms due to a
Feshbach resonance}

\author{V. A. Yurovsky}

\author{A. Ben-Reuven}

\affiliation{School of Chemistry, Tel Aviv University, 69978 Tel Aviv,
Israel}

\date{\today}
\begin{abstract}

The loss of ultracold trapped atoms in the vicinity of a
Feshbach resonance is treated as a two-stage reaction, using the
Breit-Wigner theory. The first stage is the formation of a resonant
diatomic molecule, and the second one is its deactivation by
inelastic collisions with other atoms. This model is applied to the
analysis of recent experiments on  $^{87}$Rb, leading to an estimated
value of $7\times 10^{-11}$  cm$^{3}/$s for the deactivation rate
 coefficient.

\end{abstract}


\maketitle    %JPB

The phenomenon of Feshbach resonance has received recently an
increased attention due to its application to Bose-Einstein
 condensation
(BEC) (see Ref.\ \cite{TTHK99} and references therein). Its most
outstanding effect is a drastic change of the elastic scattering
 length
as the collision energy of an atomic pair approaches the energy of a
bound level belonging to another electronic or hyperfine state. The
resonance can be tuned by applying an external magnetic field, as has
been proposed in Ref.\ \cite{Tiesinga} in order to control the BEC
properties. Applications include a controlled BEC collapse \cite{D01}
and bright solitons in BEC \cite{S02,K02}, as well as a formation of
molecular BEC \cite{TTHK99,TTCHK99,MTJ00,YBJW00,YB02}, 
an atom-molecule coherent superposition \cite{D02,KH02}, 
and an entangled atomic gas \cite{YB02}.

Another effect of the resonance is the abrupt increase in atom loss
due to inelastic collisions of the resonant molecules
\cite{TTHK99,TTCHK99,YBJW00,YBJW99}, and to the formation of 
non-condensed atoms \cite{MTJ00,YBJW00,HPW01}. 
The determination of the loss parameters is
important for an appreciation of the outcome of applications of
 Feshbach
resonances. We present here an estimate of the rate coefficient for
 the
deactivation of vibrationally excited resonant $^{87}$Rb$_{2}$
 molecules by
collisions with other Rb atoms, based on the results of recent
 experiments
\cite{M02}.

The theory presented in Refs.\ \cite{TTHK99,TTCHK99,YBJW00,YBJW99},
based on coupled Gross-Pitaevskii equations for atomic and molecular
condensates, cannot be applied to the analysis of these experiments
involving a non-condensed thermal gas. The approach used here is
 based on
the Breit-Wigner theory of resonant multichannel collisions (see e.g.
 Ref.\
\cite{MM65}), as has been proposed for the system under consideration
 by
Ref.\ \cite{MTJ00}. The reaction involving the excited
resonant molecule Rb$_{2}\left( m\right) $ includes a reversible input 
channel of formation from (and dissociation to) a pair of
 colliding atoms, 
\begin{equation}
\text{Rb + Rb }\rightleftarrows\text{ Rb}_{2}\left( m\right),
 \label{chan_e}
\end{equation}
and irreversible output channels of exoergic collisions with a third atom,
\begin{equation}
\text{Rb}_{2}\left( m\right) +\text{Rb }\rightarrow\text{
 Rb}_{2}\left( d\right) +\text{Rb}, \label{chan_d}
\end{equation}
bringing the molecule down to one of the lower-lying rovibrational
levels of the same spin state, or to levels belonging to other spin
states. (An alternative approach, presented in Refs.\ \cite{EGB99,KM02},
treats the whole process as a one-stage recombination by a three-body 
collision.)

Let us consider all atoms, for the time being, as distinguishable
particles. According to the standard theory (see Ref.\ \cite{MM65}), 
the natural resonance width $\Gamma _{e}$  associated with channel
(\ref{chan_e}) is
two times smaller than the corresponding width for the case of
indistinguishable atoms presented in Ref.\ \cite{MTJ00} (see also
Refs.\ \cite{TTHK99,Tiesinga}). It exhibits a Wigner threshold
dependence of the form
\begin{equation}
\Gamma _{e}={|a_{a}\mu |\Delta \over \hbar { } ^{2}}p ,
 \label{Gamma_e}
\end{equation}
where $a_{a}$  is the non-resonant (background) elastic scattering
length, $\mu $ is the difference of the magnetic momenta of the
 atomic pair
and the Rb$_{2}\left( m\right) $ molecule, $\Delta $ is the
 phenomenological resonance strength
(see Refs.\ \cite{TTHK99,Tiesinga}), and $p$ is the relative momentum
 of
the colliding atoms. These parameters also describe the variation of
 the
elastic scattering length $a_{\text{res}}$  as a function of the
 external magnetic
field $B$ in the vicinity of the resonance at $B=B_{0}$  as (see
 Refs.\
\cite{TTHK99,Tiesinga})
\begin{equation}
a_{\text{res}}=a_{a}\left( 1-{\Delta \over B-B{ } _{0}}\right)  .
\end{equation}
The total width $\Gamma _{d}$  associated with the deactivation
 channel
(\ref{chan_d}) can be expressed in terms of a two-body rate 
coefficient $k_{d}$, as 
\begin{equation}
\Gamma _{d}=k_{d}n 
\end{equation}
is proportional to the atomic density $n$. The rate coefficient $k_{d}$
includes the contributions of all the output deactivation channels (d)  of Eq.\ (\ref{chan_d}).

The Breit-Wigner theory leads to the following expression for the cross
section of resonance-enhanced three-body recombination (see Ref.\
\cite{MTJ00}),
\begin{equation}
\sigma ={\pi \hbar { } ^{2}\over p{ } ^{2}}{\Gamma _{e}\Gamma { }
 _{d}\over \mu ^{2}\left( B-B_{0}\right) ^{2}/\hbar ^{2}+\left(
 \Gamma _{e}+\Gamma _{d}\right) ^{2}/4}
\end{equation}
This expression does not take into account the indistinguishability
of the three participating atoms, in  which
case  the cross section should be 
$\sigma_{\text{ind}}=3!\sigma $ (see Ref.\ \cite{MM65}).

The resonant molecular state Rb$_{2}\left( m\right) $ can be formed
 whenever the
detuning from the resonance is comparable or less than $\Gamma _{e}$.
 This state
decays producing atoms with a kinetic energy spectrum of width $\hbar
 \Gamma _{e}$.
Under the conditions of the experiments \cite{M02} ($a_{a}\approx
 98.96$ atomic
units, $\mu \approx 2.8$ Bohr magnetons, $\Delta \approx 0.17$ G for
 the strongest resonance at
1007.34 G in $^{87}$Rb and a collision energy of $p^{2}/m\approx 2
 \mu $K) the width
calculated with Eq.\ (\ref{Gamma_e}) is given by $\hbar \Gamma _{e}
/k_{B}\approx 7 \mu $K, where $k_{B}$
is the Boltzmann constant. Therefore this energy is less than the trap
depth of $\approx 20 \mu $K and a spontaneous dissociation of the
 resonance molecule
(\ref{chan_e})  cannot lead to a significant loss of trapped atoms (as
opposed to the case of a BEC --- see Ref.\ \cite{MTJ00,YBJW00}). Each
deactivation event (\ref{chan_d}) leads to the simultaneous loss of
three atoms. Therefore, the loss rate for the atomic density $n\left(
 {\bf r},t\right) $ can
be written in the form,
\begin{equation}
\dot{n}\left( {\bf r},t\right) =-3{2p\over m}\sigma
 _{\text{ind}}n^{2}\left( {\bf r},t\right) =-K_{3}n^{3}\left( {\bf
 r},t\right)  , \label{ndot}
\end{equation}
where
\begin{equation}
K_{3}={36\pi \hbar ^{2}k_{d}|a_{a}\mu |\Delta \over m\left\lbrack \mu
 ^{2}\left( B-B_{0}\right) ^{2}+\hbar ^{2}\Gamma ^{2}_{e}
/4\right\rbrack } \label{K3}
\end{equation}
is the three-body loss rate coefficient. Here the partial inelastic
width $\Gamma _{d}$  is neglected in the denominator in comparison to
 $\Gamma _{e}$. Even very
close to the resonance, as long as $|B-B_{0}|>0.1$ G, the width
 $\Gamma _{e}$  may as well
be neglected, leading to an expression similar to Eq.\ (9) of
 Ref.\ \cite{YBJW99} for the loss in a BEC. 
However, the rate coefficient given
by Eq.\ (\ref{K3}) is six times larger than the corresponding rate for
a BEC. This difference, due to the effects of quantum statistics, has 
been predicted for non-resonant three-body recombination in Ref.\
\cite{KSS85}, and observed in experiments \cite{B97}.

In the case of a BEC the atomic density profile is determined by
the repulsive interaction between atoms. This interaction can be
neglected whenever its characteristic energy, proportional to the
elastic scattering length, is small compared to the kinetic energy of
atoms,
\begin{equation}
{4\pi \over m}\hbar ^{2}a_{\text{res}}n \ll k_{B}T .
\end{equation}
For the temperature $T=2 \mu $K used in the experiments \cite{M02}
 this
condition is obeyed whenever $|B-B_{0}|>0.01$ G. Therefore we can
 consider
the gas as an ideal one with the equilibrium density profile described
by the Boltzmann distribution in the trap potential.

The loss rate given by Eq.\ (\ref{ndot}) is density dependent. In
the case of an inhomogeneous trapped gas the loss processes modify the
equilibrium density profile,  leading to an atomic drift which tends
 to
compensate for this deformation. The characteristic time for this
compensation can be estimated as the trap period. In the experiments
\cite{M02} the magnetic field that brings the system close to
 resonance
has been applied during a time interval of $t=50$ ms. This time
substantially exceeds the radial trap period (the radial trap
 frequency
is $\omega _{r}/2\pi =930$ Hz), but it is less than the axial trap
 period (the axial
trap frequency is $\omega _{a}/2\pi =11$ Hz). Therefore we can
 consider the radial
density profile as an equilibrium one, described by a Boltzmann
distribution, and write out the atomic density profile as
\begin{equation}
n\left( {\bf r},t\right) ={\nu \left( z,t\right) \over \pi b{ }
 ^{2}_{r}}\exp\left( -{x^{2}+y{ } ^{2}\over b{ } ^{2}_{r}}\right)  ,
\end{equation}
where
\begin{equation}
\nu \left( z,t\right) =\int dx dy n\left( {\bf r},t\right)
\end{equation}
is a non-equilibrium axial profile and
\begin{equation}
b_{r}={1\over \omega { } _{r}}\sqrt{{2k_{B}T\over m}}
\end{equation}
is the characteristic radius of the atomic cloud.

Neglecting effects of axial atom transport, a kinetic equation for the
axial profile can be written in the form
\begin{equation}
\dot{\nu }\left( z,t\right) =-K_{1D}\nu ^{3}\left( z,t\right)
 ,\quad K_{1D}=K_{3}/\left( 3\pi ^{2}b^{4}_{r}\right)  .
 \label{nudot}
\end{equation}
The solution of Eq.\ (\ref{nudot}) relates the axial profile at time $t$
to the initial one at $t=0$ as
\begin{equation}
\nu \left( {\bf r},t\right) ={\nu \left( {\bf r},0\right) \over
 \sqrt{1+2K_{1D}\nu ^{2}\left( {\bf r},0\right) t}} .
\end{equation}
Let us suppose that at $t=0$ the atoms have a Boltzmann
distribution with the temperature $T$ and
\begin{equation}
\nu \left( {\bf r},0\right) =\nu _{0}\exp\left( -{z{ } ^{2}\over b{ }
 ^{2}_{a}}\right)  ,\qquad \nu _{0}={N{ } _{0}\over \sqrt{\pi }b{ }
 _{a}} , \label{n0}
\end{equation}
where
\begin{equation}
b_{a}={1\over \omega { } _{a}}\sqrt{{2k_{B}T\over m}}
\end{equation}
is the characteristic half-length of the atomic cloud and $N_{0}$  is
the initial number of atoms. In this case, the number of atoms
 remaining
in the trap can be expressed as
\begin{equation}
N\left( t\right) =2{N{ } _{0}\over \sqrt{\pi }} \int\limits^{\infty
 }_{0}d \zeta {\exp\left( -\zeta ^{2}\right) \over \sqrt{1+2K_{1D}\nu
 ^{2}_{0}t \exp\left( -2\zeta ^{2}\right) }} , \label{Nrem}
\end{equation}
where $\zeta =z/b_{a}$.

\begin{figure}

\includegraphics[width=3.375in]{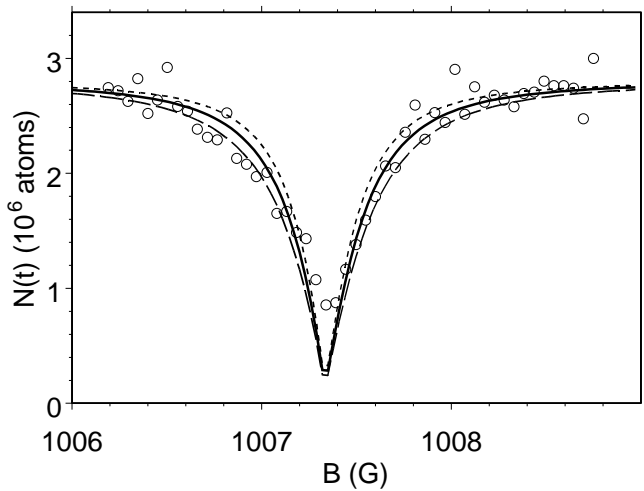}

\caption{Number of remaining atoms as a function of the magnetic
field in the vicinity of the 1007 G resonance in $^{87}$Rb calculated
 with
Eq.\ (\protect\ref{Nrem}) for three values of the deactivation rate
coefficient, $k_{d}=7\times 10^{-11}$  cm$^{3}/$s (solid line), $10^{
-10}$  cm$^{3}/$s (long-dashed
line), and $5\times 10^{-11}$  cm$^{3}/$s (short-dashed line). The
 circles represent
the experimental results of A. Marte {\it et al.} \protect\cite{M02}.}
\label{n1007}

\end{figure}

Equation (\ref{Nrem}), in combination with Eqs.\ (\ref{Gamma_e}),
(\ref{K3}), (\ref{nudot}) and (\ref{n0}), allows us to estimate the
value of $k_{d}$  by a fit to the number of remaining atoms measured
 in Ref.\
\cite{M02} for $N_{0}=2.8\times 10^{6}$. The fit produces the optimal
 value of
$k_{d}=0.7\times 10^{-10}$  cm$^{3}$/s. This value is comparable to 
corresponding estimates for Na resonances ($1.6\times 10^{-10}$  cm$^{3}/$s 
in Ref.\ \cite{YBJW00}; 
$4\times 10^{-10}$  cm$^{3}/$s and $10^{-11}$  cm$^{3}/$s
  in Ref.\
\cite{AV99} following the theory of Ref.\ \cite{TTHK99}). The
results of calculations for several values of $k_{d}$  are presented
 in Fig.\
\ref{n1007} in comparison with the experimental results of Ref.\
\cite{M02}.

The authors are most grateful to Dr. Stephan D\"{u}rr for providing
a preprint of Ref.\ \cite{M02} and clarifying details of the
experiment.

\end{document}